\documentclass[12pt]{article}
\usepackage[xdvi]{graphicx} 
\usepackage{amssymb}
\usepackage{amsmath} 
 
\begin{document}  
\newcommand{\todo}[1]{{\em \small {#1}}\marginpar{$\Longleftarrow$}}  
\newcommand{\labell}[1]{\label{#1}\qquad_{#1}} 
 
\rightline{DCPT/02/25}  
\rightline{hep-th/0204218}
\vskip 1cm 

\begin{center}
{\Large \bf Exact braneworld cosmology induced from bulk black holes}
\end{center}
\vskip 1cm
  
\renewcommand{\thefootnote}{\fnsymbol{footnote}}   \centerline{\bf
James P. Gregory\footnote{J.P.Gregory@durham.ac.uk} and Antonio
Padilla\footnote{Antonio.Padilla@durham.ac.uk}}
\vskip .5cm   \centerline{ \it Centre for Particle Theory, Department
of  Mathematical Sciences}   \centerline{\it University of Durham,
South Road, Durham DH1 3LE, U.K.}
  
\setcounter{footnote}{0} \renewcommand{\thefootnote}{\arabic{footnote}}
 

\begin{abstract}

We use a new, exact approach in calculating the energy density
measured by an observer living on a brane embedded in a charged black
hole spacetime.  We find that the bulk Weyl tensor gives rise to
non-linear terms in the energy density and pressure in the FRW
equations for the brane.  Remarkably, these take exactly the same form
as the ``unconventional'' terms found in the cosmology of branes
embedded in pure AdS, with extra matter living on the brane.  Black
hole driven cosmologies have the benefit that there is no ambiguity in
splitting the braneword energy momentum into tension and additional
matter.  We propose a new, enlarged relationship between the two
descriptions of braneworld cosmology.  We also study the exact
thermodynamics of the field theory and present a generalised
Cardy-Verlinde formula in this set up.

\end{abstract}

\newpage

\section{Introduction}

Recently there has been much interest in a holographic description of
braneworld cosmology.  Such work was inspired by Savonije and
Verlinde~\cite{Verlinde:radiation,Savonije:braneCFT} who considered a
braneworld in a Schwarzschild Anti-de Sitter (AdS) background.  They
demonstrated that the standard cosmology driven by the energy
density/pressure of an  $(n-1)$ dimensional conformal field theory
(CFT), has a dual description -- it can be regarded as being a
braneworld cosmology driven by the bulk black hole, when the brane is
near the AdS boundary. All of the details of the cosmology can be
captured by the parameters defining the background black hole
spacetime.  Furthermore, from the field theory point of view, one can
derive equations which relate the thermodynamic variables of the CFT.
As the brane evolves in the black hole background it was noticed that
these equations coincide with the cosmological evolution equations of
the braneworld at the point at which the brane crosses the black hole
horizon~\cite{Verlinde:radiation,Savonije:braneCFT}.

The notion of holography -- that all of the physics of a bulk
spacetime is encoded in a lower dimensional field theory -- is
realised in a very concise manner by the AdS/CFT
correspondence~\cite{Maldacena:adscft,Witten:adscft,Gubser:adscft}.
In this, the gravitational physics in an $n$ dimensional Anti-de
Sitter bulk spacetime is captured by an $(n-1)$ dimensional field
theory which lives on the boundary of AdS.  Witten later introduced
finite temperature into the bulk and boundary
theories~\cite{Witten:thermal}, which initiated the study of the dual
field theory for the Schwarzschild-AdS spacetime.  The study of bulk
black holes and their dual theories continued to be developed for a
wide class of spacetimes.  What will interest us here is a brane
embedded in a background of bulk Reissner-Nordstr\"om AdS black holes
(It was demonstrated in~\cite{Chamblin:charge1}, that in this case the
dual field theory is coupled to a background global current).  We
choose to study the RNAdS background as it exhibits the full
generality of many of the results discussed in earlier studies of
braneworld holography
\cite{Csaki:cosmoconst1,Csaki:cosmoconst2,Cai:chargedcardy,Biswas:stiffmatter,Myung:branecos,Cai:chargedbg,Youm:chargedcardy,Wang:branecos,Creminelli:holography}.
RNAdS has also been used to exhibit other interesting features of the
AdS/CFT correspondence which cannot be observed in the case of
non-zero temperature alone.  Firstly, because the dual field theory
has a much richer
structure~\cite{Chamblin:charge2,Evans:superfluidity}, and secondly,
because we have more control on interpreting the field theory effects
caused by the different parameters of the black hole
spacetime~\cite{Gregory:horizons}.

In braneworld holography, we normally think of the bulk gravity theory
as being dual to a CFT on the brane itself. Paradoxically, the CFT has
an ultra violet cut off, so it is actually a broken CFT
\cite{Kraus:domainwalls,Gubser:gravity}. In most studies of dual
theories on the brane, the precise nature of the dual theory is
unknown. We regard it as an abstract field theory, some of whose
properties we can derive. We should also note that early studies of
branes \cite{Randall:compactification} had fine tuned tensions that
ensured the cosmological constant on the brane vanished (critical
branes). If we avoid this fine tuning we are able to induce a non-zero
cosmological constant on the brane \cite{Karch:local} (non-critical
branes). In particular, recent observations indicating that our
universe has a small positive cosmological constant~\cite{Riess:astro,
Perlmutter:astro} suggest it is important to consider de Sitter
braneworlds. These inflationary braneworlds are naturally induced by
quantum effects of a field theory on the
brane~\cite{Hawking:newworld,Nojiri:inflation,Nojiri:brane}.

When the bulk spacetime has a non-vanishing Weyl tensor\footnote{This
can occur naturally for a hot critical braneworld due to the emission
of radiation into the bulk~\cite{Hebecker:rs2}.}, Shiromizu et al
\cite{Shiromizu:3brane} pointed out that the ``electric'' part of this
tensor appears in the Einstein's equations on the brane. This
contribution is thought of as being ``holographic'' in that it can
also be interpreted as coming from a dual theory on the brane. By
considering a braneworld observer, we can derive properties of the
dual theory and get a better understanding of the Weyl tensor
contribution. However, in all the previous literature, it has been
assumed that the brane is at large radial distance from the centre of
AdS space, i.e. near the boundary.  This allows us to assume that the
energy density of the braneworld universe is small, and the true
holographic description of an $(n-1)$ dimensional braneworld in an $n$
dimensional bulk can be understood.  However, these results are all
approximations in the sense that for a general brane evolution it is
not necessary for the brane to remain close to the boundary.  In this
paper, it is our objective to undertake a new study of the braneworld
in which we calculate the energy of the braneworld field theory
exactly, regardless of the brane's position in the bulk. We also allow
the brane tension to be arbitrary, thereby including both critical and
non-critical branes.

We find that this exact approach introduces quadratic terms in the
Friedmann Robertson Walker (FRW) equations of the braneworld
universe. Remarkably these turn out to have exactly the same form as
the quadratic terms found in the cosmology of branes embedded in a
pure AdS background, when extra matter is placed on the brane
\cite{Binetruy:branecos1,Binetruy:branecos2}. This suggests a new,
enlarged relationship between the two descriptions of braneworld
cosmology, that includes even the ``unconventional'' quadratic terms
in the FRW equation.  In the linear approximation we recover the
standard dual descriptions observed by Savonije and
Verlinde~\cite{Verlinde:radiation,Savonije:braneCFT}.  Furthermore,
unlike in \cite{Binetruy:branecos1,Binetruy:branecos2} there is no
ambiguity in splitting the braneworld energy momentum tensor into
tension and additional matter. This is because we do not have any
additional matter, only tension and a bulk Weyl tensor contribution
that we have interpreted on the brane.

The rest of the paper is organised as follows: in section 2 we derive
the equations of motion for the brane embedded in RNAdS and interpret
them as FRW equations for the brane universe. In section 3 we derive
the energy density measured by a braneworld observer exactly. We see
how the FRW equations contain the same quadratic terms found in
``unconventional' brane cosmology and suggest a new enlarged
duality. In section 4 we examine the thermodynamics of the field
theory on the brane and present a generalised (local) Cardy-Verlinde
formula that includes the exact non-linear terms. We then demonstrate
how the correspondences between the braneworld field theory and the
brane's evolution in the bulk continue to hold in this case. Finally,
section 5 contains some concluding remarks.

\section{Equations of Motion} \label{section:eom}

We will consider an $(n-1)$ dimensional brane of tension $\sigma$
sandwiched in between two $n$ dimensional black holes. In order to
show the generality of our work we are allowing the black holes to be
charged~\footnote{We note that for branes moving in a Schwarzschild
AdS bulk, there is evidence that gravity is localised on the
brane~\cite{Singh:local}.  We suspect that this still holds for a
Reissner-Nordstr\"om AdS bulk.}, although our brane will be
uncharged. Since this means that lines of flux must not converge to or
diverge from the brane, we must have black holes of equal but opposite
charge. In this case, the flux lines will pass through the brane since
one black hole will act as a source for the charge whilst the other
acts as a sink. It should be noted that although we do not have
$\mathbb{Z}_2$ symmetry across the brane for the electromagnetic
field, the geometry \bf{is} \rm $\mathbb{Z}_2$ symmetric.

We denote our two spacetimes by $\mathcal{M}^+$ and $\mathcal{M}^-$
for the positively and negatively charged black holes respectively.
Their boundaries, $\partial\mathcal{M}^+$ and $\partial\mathcal{M}^-$,
both coincide with the brane.  Our braneworld scenario is then
described by the following action:
\begin{eqnarray}
\label{eq:action}
S & = & \frac{1}{16\pi G_n}\int_{\mathcal{M}^+ + \mathcal{M}^-} d^n x
\sqrt{g}\left(R - 2\Lambda_n - F^2\right) + \frac{1}{8\pi
G_n}\int_{\partial\mathcal{M}^+ +\partial\mathcal{M}^-}
d^{n-1}x\sqrt{h} K \nonumber \\ & & + \frac{1}{4\pi
G_n}\int_{\partial\mathcal{M}^+ +\partial\mathcal{M}^-} d^{n-1}
x\sqrt{h} F^{a b} n_a A_b + \sigma\int_{brane} d^{n-1}x \sqrt{h} ,
\end{eqnarray}
where $g_{a b}$ is the bulk metric and $h_{ab}$ is the induced metric
on the brane.  $K$ is the trace of the extrinsic curvature  of the
brane, and $n_a$ is the unit normal to the brane pointing from
$\mathcal{M}^+$ to $\mathcal{M}^-$. Notice the presence of the
Hawking-Ross term in the action (\ref{eq:action}) which is necessary
for black holes with a fixed charge \cite{Hawking:ross}.

The bulk equations of motion which result from this action are given by
\begin{eqnarray}
R_{ab} - \frac{1}{2}R g_{ab} & = & -\Lambda_n g_{ab} + 2F_{ac}{F_b}^c
- \frac{1}{2}g_{ab} F^2 \\  \partial_a\left(\sqrt{g} F^{ab}\right) & =
& 0
\end{eqnarray}
These admit the following 2 parameter family of electrically charged
black hole solutions for the bulk metric
\begin{equation} \label{eqn:metric}
{ds_n}^2 = -h(Z)dt^2 + \frac{dZ^2}{h(Z)} + Z^2 d\Omega_{n-2},
\end{equation}
in which
\begin{equation} \label{eqn:h(Z)}
h(Z) = {k_n}^2 Z^2 + 1 - \frac{c}{Z^{n-3}} + \frac{q^2}{Z^{2n-6}},
\end{equation}
and the electromagnetic field strength
\begin{equation} \label{eqn:fieldstrength}
F = dA \quad \textrm{where} \quad A =
\left(-\frac{1}{\kappa}\frac{q}{Z^{n-3}} + \Phi\right) dt \quad
\textrm{and} \quad \kappa = \sqrt{\frac{2(n-3)}{n-2}}.
\end{equation}
Note that $d\Omega_{n-2}$ is the metric on a unit $(n-2)$
sphere. $k_n$ is related to the bulk cosmological constant by
$\Lambda_n=-\frac{1}{2}(n-1)(n-2)k_n^2$, whereas $c$ and $q$ are
constants of integration. If $q$ is set to zero in this solution, we
regain the AdS-Schwarzschild solution in which $c$ introduces a black
hole mass.  The presence of $q$ introduces black hole charge for which
$\Phi$ is an electrostatic potential difference.  In this general
metric, $h(Z)$ has two zeros, the larger of which, $Z_H$, represents
the event horizon of the black hole.

Here, charge is a localised quantity. It can be evaluated from a
surface integral on any closed shell wrapping the black hole (Gauss'
Law). In $\mathcal{M}^{\pm}$ the total charge is
\begin{equation}
Q=\pm \frac{(n-2)\kappa \Omega_{n-2}}{8\pi G_n} q,
\end{equation}  
where $\Omega_{n-2}$ is the volume of a unit $(n-2)$ sphere. Note also
that the ADM mass \cite{Hawking:adsblackholes} of each black hole is
given by
\begin{equation}
M=\frac{(n-2)\Omega_{n-2}c}{16 \pi G_n}.
\end{equation}

Let us now consider the dynamics of our brane embedded in this
background of charged black holes. We use $\tau$ to parametrise the
brane so that it is given by the section $({\bf{x} \rm}^{\mu},
t(\tau), Z(\tau))$ of the bulk metric.  The Israel equations for the
jump in extrinsic curvature across the brane give the brane's
equations of motion. One might suspect that the presence  of the
Hawking-Ross term in the action will affect the form of these
equations. However, since the charge on the black holes is fixed, the
flux across the brane does not vary and the Israel equations take
their usual form
\begin{equation}
2K_{ab} - 2K h_{ab} = -8\pi G_n \sigma h_{ab},
\end{equation}
where
\begin{equation}
K_{ab} = h_{a}^c h_{b}^d \nabla_{\left(c\right.}n_{\left.d\right)}
\quad \textrm{and} \quad n_a=({\bf 0}, -\dot{Z},\dot{t}).
\end{equation}
Here we use an overdot to denote differentiation with respect to
$\tau$.  Note that $n_a$ is the \bf{unit} \rm normal, so we have the
condition

\begin{equation} \label{eqn:condition}
-h(Z)\dot{t}^2+\frac{\dot{Z}^2}{h(Z)}=-1
\end{equation}
The resulting equations of motion for the brane are:
\begin{subequations}
\begin{eqnarray}
\dot{Z}^2 & = & a Z^2 - 1 + \frac{c}{Z^{n-3}} - \frac{q^2}{Z^{2n-6}}
\label{eqn:evolution:a}\\ \ddot{Z} & = & a Z -
\left(\frac{n-3}{2}\right)\frac{c}{Z^{n-2}} +
(n-3)\frac{q^2}{Z^{2n-5}} \label{eqn:evolution:b}\\ \dot{t} & = &
\frac{\sigma_n Z}{h(Z)} \label{eqn:evolution:c}
\end{eqnarray}
\end{subequations}
where $a=\sigma_n^2 - k_n^2$ and $\sigma_n = \frac{4\pi G_n
\sigma}{n-2}$. This analysis has been presented in more detail, at
least for uncharged black holes, in \cite{Bowcock:branecos,
Petkou:brane}.

We shall now examine the cosmology of our braneworld in this
background. The induced metric is given by the following
\begin{equation}
ds_{n-1}^2 = -d\tau^2 + Z(\tau)^2 d\Omega_{n-2}.
\end{equation}
We see that we may interpret $Z(\tau)$ as corresponding to the scale
factor in an  expanding/contracting universe and that equations
(\ref{eqn:evolution:a}) and (\ref{eqn:evolution:b}) should be regarded
as giving rise to the Friedmann equations of our braneworld. Indeed,
if we define the Hubble parameter as $H=\frac{\dot{Z}}{Z}$, we arrive
at the following equations for the cosmological evolution of the brane
\begin{subequations}  \label{eqn:FRW}
\begin{eqnarray}
H^2 & = & a - \frac{1}{Z^2} + \frac{c}{Z^{n-1}} - \frac{q^2}{Z^{2n-4}}
\label{eqn:FRW:a} \\
\dot{H} & = & \frac{1}{Z^2} -
\left(\frac{n-1}{2}\right)\frac{c}{Z^{n-1}}
+(n-2)\frac{q^2}{Z^{2n-4}}. \label{eqn:FRW:b}
\end{eqnarray}
\end{subequations}
Let us examine these equations in more detail. Equation
(\ref{eqn:FRW:a}) contains the cosmological constant term $a$. For
$a=0$ we have a critical wall with vanishing cosmological
constant. For $a > 0 /a<0$ we have super/subcritical walls that
correspond to asymptotically de Sitter/anti-de Sitter spacetimes. Note
that for subcritical and critical walls, $Z$ has a maximum and minimum
value. For supercritical walls, we have two possibilities: either $Z$
is bounded above and below or it is only bounded below and may stretch
out to infinity.

However, our real interest in equations (\ref{eqn:FRW:a}) and
(\ref{eqn:FRW:b}), lies in understanding the $c$ and $q^2$ terms. As
discussed in \cite{Savonije:braneCFT, Padilla:CFT}, we can interpret
the contribution of the $c$ term to the cosmological evolution in two
different ways. On the one hand, the evolution  is driven, in part, by
the masses of the bulk black holes, as is evident in equations
(\ref{eqn:FRW:a}) and (\ref{eqn:FRW:b}).  On the other hand, we can
ignore the bulk and describe the evolution as  being driven by the
energy density and pressure of radiation in a dual field theory living
on the brane. At least for critical walls, it was showed by Biswas and
Mukherji~\cite{Biswas:stiffmatter} that we could interpret the $q^2$
terms as corresponding to stiff matter in the dual theory. These
discussions were motivated by ``holography'' in the sense that the
duality related an $n$ dimensional bulk theory to an $(n-1)$
dimensional field theory on the brane. In the spirit of AdS/CFT, the
brane was close to the AdS boundary. In the next section we will not
make that assumption. We will nevertheless discover an interesting
alternative picture, although it will be ``unconventional'', at least
in terms of cosmology!

\section{Energy on the brane}

Consider an observer living on the brane. He measures time using the
braneworld coordinate, $\tau$, rather than the bulk time coordinate,
$t$. This of course affects his measurement of the energy density. In
previous literature, the energy on the brane has been obtained by
scaling the energy of the bulk accordingly. For example, in
\cite{Padilla:CFT}, a detailed calculation using Euclidean gravity
techniques yielded a bulk energy, $E_{bulk} \approx
2M\left(\frac{k_n^2}{\sigma_n^2}\right)$, where $M$ is the ADM mass of
the black holes. This was then scaled using $\dot{t}$ to give the
energy on the brane. In the limit of large $Z(\tau)$, $\dot{t} \approx
\frac{\sigma_n}{k_n^2Z}$ so that a braneworld observer measures the
energy to be $E \approx \frac{2M}{\sigma_n Z}$. This analysis enabled
us to have the dual picture described in the previous section, even
for non-critical branes.

While the results obtained from these methods are interesting, they
are approximations. Indeed, the Euclidean analysis of
\cite{Padilla:CFT} is only valid in certain limits, imposed because we
require time translational symmetry to Wick rotate to Euclidean
signature. We should note that the deviation from the expected bulk
energy, $2M$, is not so startling as we might originally think. The
limits of the analysis impose that as the brane approaches the AdS
boundary, $\sigma_n \to k_n$. Therefore, if the brane actually strikes
the AdS boundary, it has to be critical and we recover what we might
have na\"\i vely expected: the bulk energy is given by the sum of the
ADM masses.

In this paper, we will bypass all of these approximations and
limitations by ignoring the bulk energy and calculating the energy on
the brane directly. We will use the techniques of
\cite{Hawking:hamiltonian} to evaluate the gravitational energy using
$\tau$ as our chosen time coordinate. Happily we will not need to Wick
rotate to Euclidean signature, enabling us to exactly calculate the
energy density on the brane, even at smaller values of $Z(\tau)$.

We begin by focusing on the contribution from the positively charged
black hole spacetime, $\mathcal{M}^+$ and its boundary,
$\partial\mathcal{M}^+$.  This boundary of course coincides with the
brane. Consider the timelike vector field defined on $\partial
\mathcal{M}^+$

\begin{equation} \label{eqn:tau}
\tau^a=(\mathbf{0}, \dot{t}, \dot{Z}).
\end{equation}
This maps the boundary/brane onto itself, and satisfies $\tau^a
\nabla_a \tau=1$.  In principle we can extend the definition of
$\tau^a$ into the bulk, stating only that it approaches the form given
by equation (\ref{eqn:tau}) as it nears the brane. We now introduce a
family of spacelike surfaces, $\Sigma_\tau$, labelled by $\tau$ that
are always normal to $\tau^a$. This family provide a slicing of the
spacetime, $\mathcal{M}^+$ and each slice meets the brane
orthogonally. As usual we decompose $\tau^a$ into the lapse function
and shift vector, $\tau^a=N r^a +N^a$, where $r^a$ is the unit normal
to $\Sigma_\tau$. However, when we lie on the brane, $\tau^a$ \bf{is}
\rm the unit normal to $\Sigma_\tau$, because there we have the
condition (\ref{eqn:condition}). Therefore, on $\partial
\mathcal{M}^+$, the lapse function, $N=1$ and the shift vector,
$N^a=0$. Before we consider whether or not we need to subtract off a
background energy, let us first state that the relevant part of the
action at this stage of our analysis is the following:

\begin{equation}
I^+ =  \frac{1}{16\pi G_n}\int_{\mathcal{M}^+} R - 2\Lambda_n - F^2 +
\frac{1}{8\pi G_n}\int_{\partial \mathcal{M}^+}  K  + \frac{1}{4\pi
G_n}\int_{\partial\mathcal{M}^+} F_{a b} n^a A^b.
\end{equation}
As stated earlier, we do not include any contribution from
$\mathcal{M}^-$ or $\partial \mathcal{M}^-$, nor do we include the
term involving the brane tension. This is because we want to calculate
the gravitational Hamiltonian, without the extra contribution of a
source. The brane tension has already been included in the analysis as
a cosmological constant term, and it would be wrong to double count.

Given the slicing $\Sigma_\tau$, the Hamiltonian that we derive from
$I^+$ is given by

\begin{eqnarray} \label{eqn:hamiltonian}
H^+ &=& \frac{1}{8 \pi G_n} \int_{\Sigma_{\tau}} N\mathcal{H} + N^a
\mathcal{H}_a - 2NA_{\tau} \nabla_a E^a \nonumber \\ & & -\frac{1}{8
\pi G_n} \int_{S_{\tau}} N \Theta + N^a p_{ab}n^b - 2NA_{\tau}n_a E^a
+ 2NF^{ab}n_aA_b
\end{eqnarray}
where $\mathcal{H}$ and $\mathcal{H}_a$ are the Hamiltonian and the
momentum constraints respectively. $p^{ab}$ is the canonical momentum
conjugate to the induced metric on $\Sigma_{\tau}$ and $E^a$ is the
momentum conjugate to $A_a$.  The surface $S_\tau$ is the intersection
of $\Sigma_{\tau}$ and the brane, while $\Theta$ is the trace of the
extrinsic curvature of $S_{\tau}$ in $\Sigma_\tau$.

Note that the momentum $E^a=F^{a \tau}$. In particular, $E^{\tau}=0$
and we regard $A_{\tau}$ as an ignorable coordinate. We will now
evaluate this Hamiltonian for the RNAdS spacetime described by
equations (\ref{eqn:metric}), (\ref{eqn:h(Z)}) and
(\ref{eqn:fieldstrength}). Each of the constraints vanish because this
is a solution to the equations of motion.

\begin{equation}
\mathcal{H}=\mathcal{H}_a=\nabla_a E^a=0.
\end{equation}
The last constraint is of course Gauss' Law. When evaluated on the
surface $S_{\tau}$,  the potential,
$A=\left(-\frac{1}{\kappa}\frac{q}{Z(\tau)^{n-3}} + \Phi\right)
\dot{t} \ d\tau$. The important thing here is that it  only has
components in the $\tau$ direction. This ensures that the last two
terms in the Hamiltonian cancel one another. Since $N=1$ and $N^a=0$
on $S_{\tau}$, it only remains to evaluate the extrinsic curvature
$\Theta$. If $\gamma_{ab}$ is the induced metric on $S_{\tau}$, it is
easy to show that

\begin{equation}
\Theta=\Theta_{ab} \gamma^{ab}=K_{ab} \gamma^{ab} = (n-2)\frac{h(Z)
\dot{t}}{Z}.
\end{equation}
The energy is then evaluated as
\begin{equation}
\mathcal{E}= -\frac{1}{8 \pi G_n} \int_{S_{\tau}} (n-2)\frac{h(Z)
\dot{t}}{Z}.
\end{equation}
We will now address the issue of background energy. This is usually
necessary to cancel divergences in the Hamiltonian.  In our case, the
brane cuts off the spacetime. If the brane does not stretch to the AdS
boundary there will not be any divergences that need to be
cancelled. However it is important to define a zero energy
solution. In this work we will choose pure AdS space. This is because
the FRW equations for a brane embedded in pure AdS space would include
all but the holographic terms that appear in equations
(\ref{eqn:FRW:a}) and (\ref{eqn:FRW:b}). These are the terms we are
trying to interpret with this analysis.

We will denote the background spacetime by $\mathcal{M}_0$.  We have
chosen this to be pure AdS space cut off at a surface $\partial
\mathcal{M}_0$ whose geometry is the same as our brane. As is
described in \cite{Padilla:CFT} this means we have the bulk metric
given by

\begin{equation} \label{eqn:AdSmetric}
{ds_n}^2 = -h_{AdS}(Z)dT^2 + \frac{dZ^2}{h_{AdS}(Z)} + Z^2
d\Omega_{n-2},
\end{equation}
in which
\begin{equation} \label{eqn:h_AdS(Z)}
h_{AdS}(Z) = {k_n}^2 Z^2 + 1.
\end{equation}
There is of course no electromagnetic field. The surface  $\partial
\mathcal{M}_0$ is described by the section $(\mathbf{x}^{\mu},
T(\tau), Z(\tau))$ of the bulk spacetime. In order that this surface
has the same geometry as our brane we impose the condition

\begin{equation} \label{eqn:AdScondition}
-h_{AdS}(Z)\dot{T}^2+\frac{\dot{Z}^2}{h_{AdS}(Z)}=-1
\end{equation} 
which is analogous to the condition given in equation
(\ref{eqn:condition}).

We now repeat the above evaluation of the Hamiltonian for the
background spacetime. This gives the following value for the
background energy

\begin{equation}
\mathcal{E}_0= -\frac{1}{8 \pi G_n} \int_{S_{\tau}}
(n-2)\frac{h_{AdS}(Z) \dot{T}}{Z}.
\end{equation}
Making use of equations (\ref{eqn:evolution:a}),
(\ref{eqn:evolution:c}) and (\ref{eqn:AdScondition}) we find that the
energy of $\mathcal{M}^+$ above the background $\mathcal{M}_0$ is
given by

\begin{equation}
E_+=\mathcal{E}-\mathcal{E}_0=\frac{(n-2)}{8 \pi G_n} \int_{S_{\tau}}
\sqrt{\sigma_n^2-\frac{\Delta h}{Z^2}} -\sigma_n
\end{equation}
where
\begin{equation}
\Delta h=h(Z)-h_{AdS}(Z)=-\frac{c}{Z^{n-3}}+\frac{q^2}{Z^{2n-6}}.
\end{equation}
In this relation $\Delta h$ is negative everywhere outside of the
black hole horizon and so it is clear that $E_+$ is positive.  We now
turn our attention to the contribution to the energy from
$\mathcal{M}^-$. Since the derivation of $E_+$ saw the cancellation of
the last two terms in the Hamiltonian (\ref{eqn:hamiltonian}) we note
that the result is purely geometrical. Even though $\mathcal{M}^+$ and
$\mathcal{M}^-$ have opposite charge, they have the same geometry and
so $E_+=E_-$. We deduce then that the total energy

\begin{equation} \label{eqn:energy}
E=E_++E_-= \frac{(n-2)}{4 \pi G_n} \int_{S_{\tau}}
\sqrt{\sigma_n^2-\frac{\Delta h}{Z^2}} -\sigma_n
\end{equation}
Since the spatial volume of the braneworld
$V=\int_{S_{\tau}}=\Omega_{n-2}Z^{n-2}$, we arrive at the following
expression for the energy density measured by an observer living on
the brane

\begin{equation} \label{eqn:rho}
\rho= \frac{(n-2)\sigma_n}{4 \pi G_n} \left( \sqrt{1-\frac{\Delta
h}{\sigma_n^2Z^2}} -1 \right)
\end{equation}
where we have pulled out a factor of $\sigma_n$. If we insert this
expression back into the first Friedmann equation (\ref{eqn:FRW:a}) we
find that

\begin{equation} \label{eqn:FRWft:a}
H^2= a - \frac{1}{Z^2} + \frac{8 \pi G_n \sigma_n}{n-2} \rho + \left (
\frac{4 \pi G_n}{n-2} \right)^2 \rho^2
\end{equation}
Although we will leave a more detailed analysis of the pressure, $p$,
until section \ref{section:thermodynamics}, we can make use of the
Conservation of Energy equation
\begin{equation}
\dot{\rho}=-(n-2)H(\rho+p)
\end{equation}
to derive the second Friedmann equation
\begin{equation} \label{eqn:FRWft:b}
\dot{H}=\frac{1}{Z^2}-4 \pi G_n \sigma_n (\rho+p) -(n-2)\left (
\frac{4 \pi G_n}{n-2} \right)^2\rho  (\rho+p)
\end{equation}

These are clearly not the standard Friedmann equations for an $(n-1)$
dimensional universe with energy density $\rho$ and pressure
$p$. However, we should not expect them to be. We have not made any
approximations in arriving at these results so it is possible that we
would see non-linear terms. What is exciting is that the quadratic
terms we see here have exactly the same form as the unconventional
terms that were originally noticed by
\cite{Binetruy:branecos1,Binetruy:branecos2} in the study of brane
cosmologies. In that case, one places extra matter on the brane to
discover this unconventional cosmology. We have no extra matter on the
brane but by including a bulk black hole, we get exactly the same type
of cosmology. Clearly there is an alternative description.

We also note that in \cite{Binetruy:branecos1,Binetruy:branecos2}, the
energy momentum tensor on the brane is split between tension and
additional matter in an arbitrary way.  In our analysis the tension is
the \bf {only} \rm explicit source of energy momentum on the brane so
there is no split required.  With this in mind we are able to
interpret each term in the FRW equations more confidently, in
particular, the cosmological constant term.  Furthermore, we have not
yet  made any assumptions on the form of the braneworld Newton's
constant.

Finally, we see that for small $\rho$ and $p$, we can neglect the
$\rho^2$ and $\rho p$  terms and recover the standard Friedmann
equations for an $(n-1)$ dimensional universe
\begin{eqnarray}
H^2 &=& a - \frac{1}{Z^2} + \frac{16 \pi G_{n-1}}{(n-2)(n-3)} \rho
\label{eqn:linearFRW:a} \\
\dot{H} &=& \frac{1}{Z^2}- \frac{8 \pi G_{n-1}}{(n-3)} (\rho+p)
\label{eqn:linearFRW:b}
\end{eqnarray}
Here we have taken the $(n-1)$ dimensional Newton's Constant to be

\begin{equation} \label{eqn:newton}
G_{n-1}=\frac{(n-3)}{2}G_n \sigma_n
\end{equation}
as is suggested by
\cite{Randall:compactification,Shiromizu:3brane,Uchida:radion,Gregory:nested,Gregory:instantons,
Singh:local}. We see, then, how the relationships noticed in previous
studies of braneworld holography are just an approximation of the
relationship described here.

\section{Thermodynamics on the brane} \label{section:thermodynamics}

In \cite{Savonije:braneCFT} it was noticed that there was a remarkable
relationship between the thermodynamics of a field theory on the brane
and its gravitational dynamics in the AdS bulk. That work was of
course based on the assumption that the brane reached close to the AdS
boundary. Our work is not limited by these approximations and we shall
henceforth attempt to generalise those results to this exact setting.

We shall assume that our field theory on the brane is in thermodynamic
equilibrium. We would therefore expect it to satisfy the first law of
thermodynamics
\begin{equation} \label{eqn:firstlaw}
T\mathrm{d}S = \mathrm{d}E - \Phi \mathrm{d}Q + p\mathrm{d}V.
\end{equation}

Notice the presence of the chemical potential $\Phi$, which is
conjugate to the R charge, $Q$ of the field theory. This is typical of
theories that are dual to RNAdS in the bulk \cite{Cvetic:rcharge}.

In reality, this first law arises from the contributions of both of
our black holes, $\mathcal{M}^+$ and $\mathcal{M}^-$.  Written in
terms of the field theory quantities which are derived from each of
these black holes, the first law should actually take the extended form
\begin{equation}
T_+\mathrm{d}S_+ + T_-\mathrm{d}S_- = \mathrm{d}E_+ + \mathrm{d}E_- -
\Phi_+ \mathrm{d}Q_+ - \Phi_- \mathrm{d}Q_- + (p_+ + p_-)\mathrm{d}V.
\end{equation}
We noted in section~\ref{section:eom} that $\mathcal{M}^+$ and
$\mathcal{M}^-$ have the same geometry. Field theory quantities that
only see this geometry, as opposed to the difference in charge, will
therefore be the same whether they are derived from  $\mathcal{M}^+$
or $\mathcal{M}^-$. We deduce then that we should derive a single
temperature, $T$, entropy, $S$, energy, $E$ and pressure, $p$ where

\begin{eqnarray}
T &=& T_+ \ \ = \  T_- \nonumber \\ S &=& 2S_+ \ = \ 2S_- \nonumber \\
E &=& 2E_+ \ = \ 2E_-  \nonumber \\ p &=& 2p_+ \ = \ 2p_- \nonumber.
\end{eqnarray}

However we need to be more careful in discussing the chemical
potential and the charge because these are aware of the different sign
in the charges in  $\mathcal{M}^+$ and $\mathcal{M}^-$.  Since the
charges are equal and opposite we define
\begin{eqnarray}
\Phi &=& \Phi_+ \ \ = \ -\Phi_- \nonumber \\ Q &=& 2Q_+ = -2Q_-
\nonumber
\end{eqnarray}
Given these relations, we do indeed recover a simplified first law of
the form~(\ref{eqn:firstlaw}).

To proceed further we will have to assume that the entropy and charge
of our field theory is given exactly by the entropy and charge of the
black holes.  Taking into account the doubling up from the two black
holes we find that

\begin{equation} \label{eqn:S}
S=\frac{\Omega_{n-2}Z_H^{n-2}}{2G_n} \quad \textrm{and} \quad
Q=\frac{(n-2)\kappa \Omega_{n-2}}{4\pi G_n} q.
\end{equation}
Recall that the energy $E$ is given by
\begin{equation}
E = \frac{(n-2)\sigma_n\Omega_{n-2}Z^{n-2}}{4\pi G_n}\bigg(\xi(Z) -
1\bigg).
\end{equation}
where we have taken out a factor of $\sigma_n$ in equation
(\ref{eqn:energy}) and used the simplifying notation
\begin{equation}
\xi(Z) = \sqrt{1-\frac{\Delta h}{\sigma_n^2 Z^2}}.
\end{equation}

The remaining thermodynamic variables, $p$,  $T$ and  $\Phi$, can now
be obtained by making use of the first law~(\ref{eqn:firstlaw}).
\begin{eqnarray}
p = -\left(\frac{\partial E}{\partial V}\right)_{S,Q} & = &  -\rho
+\frac{1}{8\pi G_n\sigma_n \xi(Z)} \left[\frac{(n-1)c}{Z^{n-1}} -
\frac{2(n-2)q^2}{Z^{2n-4}}\right] ,\\ T = \left(\frac{\partial
E}{\partial S}\right)_{Q,V} & = &
\frac{1}{\xi(Z)}\frac{T_{BH}}{\sigma_n Z}, \\ \Phi =
\left(\frac{\partial E}{\partial Q}\right)_{S,V} & = & \frac{1}{\kappa
\sigma_n\xi(Z)Z} \left(\frac{q}{Z_H^{n-3}}-\frac{q}{Z^{n-3}}\right),
\end{eqnarray}
where $T_{BH}=\frac{h'(Z_H)}{4\pi}$ is the temperature of the black
holes.  Given the value of $p$ that we have derived above, it is a
useful check of the consistency of these results, to observe that
equations~(\ref{eqn:FRWft:b}) and~(\ref{eqn:FRW:b}) are indeed the
same. Notice also that the chemical potential, $\Phi$, vanishes at the
horizon.

In earlier studies of such braneworlds, the nonlinear terms in $\rho$
were not taken into account due to various approximations which were
made during the calculations.  It is therefore far from trivial to
state that any further results connecting the bulk spacetime physics
and the field theory thermodynamics should continue to hold in our
scenario.  We therefore proceed to consider these connections between
the FRW equations and the field theory variables, including all such
non-linear terms, to observe how the connection is affected in their
presence.

Our first task is to rewrite the first law of thermodynamics in terms
of densities and we find that
\begin{equation}
T\mathrm{d}s = \mathrm{d}\rho - \Phi\mathrm{d}\rho_Q - \gamma \,
\mathrm{d}\left(\frac{1}{Z^2}\right),
\end{equation}
where
\begin{equation} \label{eqn:cardy}
\gamma = \frac{(n-2)Z^2}{2}(\rho + p - \Phi\rho_Q - Ts).
\end{equation}
and the densities $s=\frac{S}{V}$ and $\rho_Q=\frac{Q}{V}$. This
equation defines $\gamma$ which is the variation of $\rho$ with
respect to the spatial curvature $1/Z^2$. It therefore represents the
geometrical Casimir part of the energy density.  Using the values we
have derived for our field theory thermodynamic variables, we find that
\begin{equation}
\gamma = \frac{n-2}{8\pi G_n\sigma_n\xi(Z)}\frac{Z_H^{n-3}}{Z^{n-3}}.
\end{equation}
We are now ready to present a generalised (local) Cardy-Verlinde
formula for the entropy density, $s$, in terms of $\gamma$ and the
thermodynamic variables
\begin{eqnarray} \label{eqn:scardy}
s & = & \frac{4\pi\sigma_n}{(n-2)k_n}   \Bigg[  \left(1+\frac{4\pi
   G_n\rho}{(n-2)\sigma_n}\right)^2
   \gamma\left(\rho-\frac{1}{2}\Phi\rho_Q-\frac{\gamma}{Z^2}\right)
   \phantom{1111111111111}\Bigg. \\ &&
   \Bigg. \phantom{1111111111111111111111111}  -\frac{2\pi
   G_n}{(n-2)\sigma_n}\rho^2 \gamma \left(1+\frac{4\pi G_n
   \rho}{(n-2)\sigma_n}\right) \Bigg]^{1/2}. \nonumber
\end{eqnarray}
In the limit of small $\rho$, we can ignore the quadratic terms and
this reduces to the formula found in
\cite{Savonije:braneCFT,Cai:chargedcardy,Biswas:stiffmatter,Medved:holography}
\begin{equation} \label{eqn:cardy:linear}
s =\frac{4\pi\sigma_n}{(n-2)k_n}
\sqrt{\gamma\left(\rho-\frac{1}{2}\Phi\rho_Q-\frac{\gamma}{Z^2}\right)}
\end{equation}

In \cite{Savonije:braneCFT}, it was noticed that, when a critical
brane $(\sigma_n=k_n)$ crosses the black hole horizon, the Hubble
Parameter is given by $H^2=k_n^2$. At this instance, the entropy
density can be written as
\begin{equation} \label{eqn:entropy}
s=\frac{H}{2G_n k_n}=\frac{(n-3)H}{4G_{n-1}}
\end{equation}
where we have made use of equation (\ref{eqn:newton}) when
$\sigma_n=k_n$. Notice that this corresponds to the Hubble entropy
described in \cite{Verlinde:radiation}.  The key observation is that,
at the horizon, equation (\ref{eqn:cardy:linear}) then reduces to the
linear form of the first Friedmann equation (\ref{eqn:linearFRW:a}).
Confining ourselves to the case of critical branes, this result
generalises to the non-linear case when we evaluate equation
(\ref{eqn:scardy}) at the horizon, since we then recover the exact
form of the Friedmann equation (\ref{eqn:FRWft:a}).

For non-critical branes $(\sigma_n \neq k_n)$ at the horizon, the
connection between entropy density and the Hubble parameter is not a
linear one.  We instead find that $H^2=\sigma_n^2=k_n^2+a$, therefore
using equation~(\ref{eqn:S}), the entropy density on non-critical
branes should be rewritten as
\begin{equation} \label{eqn:s2}
s=\frac{\sqrt{H^2-a}}{2G_n k_n}=\left(\frac{\sigma_n}{k_n}
\right)\frac{(n-3)\sqrt{H^2-a}}{4 G_{n-1}}
\end{equation}
where we have once again made use of equation
(\ref{eqn:newton}). Unlike in the critical case, it is not clear how
we should interpret this entropy formula. Nevertheless we note that
when equations (\ref{eqn:cardy:linear}) and (\ref{eqn:scardy}) are
evaluated at the horizon they reproduce the linear
(\ref{eqn:linearFRW:a}) and non-linear (\ref{eqn:FRWft:a}) Friedmann
equations respectively.

We also present the generalised form of the temperature at the horizon
crossing
\begin{equation} \label{eqn:T2}
T=-\frac{k_n\dot{H}}{2\pi\sigma_n\sqrt{H^2-a}} \left(1+\frac{4\pi
 G_n\rho}{(n-2)\sigma_n}\right)^{-1}.
\end{equation}
Notice that this reduces to the formula quoted in
\cite{Savonije:braneCFT} when we ignore quadratic terms and set
$\sigma_n=k_n$ for the critical brane. Using this generalised form, we
see that if we evaluate equation (\ref{eqn:cardy}) at the horizon, we
reproduce the second Friedmann equation (\ref{eqn:FRWft:b}), even in
the exact, non-linear case.

\section{Discussion}

We have noticed a remarkable new relationship for braneworld cosmology
that includes the unconventional terms first spotted in
\cite{Binetruy:branecos1,Binetruy:branecos2}. Consider the approach of
\cite{Binetruy:branecos1,Binetruy:branecos2} and embed a brane in pure
AdS space with additional matter placed on the brane. The braneworld
observer sees a different cosmology to what we might expect. The
Friedmann equations include quadratic terms in the energy density and
pressure of the additional matter. In our work, we see that we obtain
exactly the same unconventional terms when one considers a brane
embedded in between two black holes, with no extra matter being placed
on the brane. In previous studies of branes in black hole backgrounds,
the cosmological evolution has been described in two different
ways. As above, it may be regarded as being driven by the bulk black
hole or by a field theory living on the brane. However, these results
assume that the brane is close to the AdS boundary. We have used
techniques that do not require us to make such approximations in our
calculations. It seems that the ``holographic'' relationships of
\cite{Savonije:braneCFT} are just an approximation of the
``unconventional'' relationship we have noticed.

Another appealing feature of our analysis is that it is free of
ambiguities. In~\cite{Binetruy:branecos1,Binetruy:branecos2} it is not
clear why we should split the brane energy momentum tensor into
tension and additional matter in the way that we do. In our approach
there is no additional matter, only tension, so there is no ambiguous
split. We find that our exact interpretation of the bulk Weyl tensor
suggests that the FRW equations should indeed take the form given in
\cite{Binetruy:branecos1,Binetruy:branecos2}. However, by comparing
our values for $\rho$ and $p$ we also see that we have a highly
non-trivial equation of state. It only takes a recognisable form as
the brane approaches the AdS boundary. This is expected because by
allowing the brane to probe deep into the bulk we are inserting a
significant cut-off in the braneworld field theory. The approach of
\cite{Binetruy:branecos1,Binetruy:branecos2} has the benefit that it
allows for an arbitrary linear equation of state.

Given our exact work, we have also been able to present a generalised
Cardy-Verlinde formula for the field theory on the brane. We should
note that we made various assumptions (e.g. entropy on brane is
entropy of bulk) that perhaps require further justification -- Our
results are as given but it would be worth taking a closer look at how
a braneworld observer should really see entropy in this exact
setting. Nevertheless, we find that just as in
\cite{Savonije:braneCFT}, the Friedmann equations are reproduced when
we evaluate the relevant thermodynamic formulas at the horizon,
including all non-linear terms. We should note that we do not know how
we should interpret the expression for the entropy (\ref{eqn:s2}) when
there is a cosmological constant induced on the brane.

\vskip .5in

\centerline{\bf Acknowledgements}
\medskip

We would like to thank Breacher, Taff, Lord Lovis, Saffers, and the
Page Boy for lively discussions. In particular, we would like to thank
Ruth Gregory, Clifford Johnson, and Simon Ross for their useful
comments. JPG is supported in part by EPSRC studentship 9980045X.  AP
was funded by PPARC.

\bibliographystyle{utphys}

\bibliography{exactcosmo}
 
\end{document}